\documentclass[12pt]{article}

\usepackage{sbc-template}

\usepackage{graphicx,url}

\usepackage[latin1]{inputenc}  

\usepackage{amsmath}
\usepackage{array, amsfonts}
\usepackage{amsthm}
\usepackage{amssymb}
\usepackage{enumerate}
\usepackage{lineno}
\usepackage{enumitem}
\usepackage{mathrsfs}
\usepackage{color}

\definecolor{darkgreen}{rgb}{0,0.5,0}

\newcommand{\comment}[1]

\title{Expected Emergence of Algorithmic Information from a Lower Bound for Stationary Prevalence}

\author{ Felipe S. Abrah\~{a}o, Klaus Wehmuth, Artur Ziviani}

\address{National Laboratory for Scientific Computing (LNCC) \\ Av. Get\'{u}lio Vargas, 333, Quitandinha -- CEP 25651-075 -- Petr\'{o}polis, RJ -- Brazil
	\email{fsa@lncc.br, 
		klaus@lncc.br, 
		ziviani@lncc.br}
}

\theoremstyle{definition}

\theoremstyle{remark}

\theoremstyle{remark}

\theoremstyle{definition}

\theoremstyle{remark}

\theoremstyle{remark}

\begin{document} 
	
	\maketitle
	
	\begin{abstract}
		We study emergent information in populations of randomly generated networked computable systems that follow a Susceptible-Infected-Susceptible contagion (or infection) model of imitation of the fittest neighbor. These networks have a scale-free degree distribution in the form of a power-law following the Barab\'{a}si-Albert model.  We show that there is a lower bound for the stationary prevalence (or average density of infected nodes) that triggers an unlimited increase of the expected emergent algorithmic complexity (or information) of a node as the population size grows.
	\end{abstract}

	\section{Introduction}
	The general scope of this work involves complex systems, complex networks, information theory, and computability theory. In particular, we aim at studying the general problem of emergence of complexity or information when complex systems are networked compared with when they are isolated. This issue has a pervasive importance in the literature about complex systems with applications on investigating systemic properties of biological, economical, or social systems. Following a theoretical approach on this subject, we present a study on the emergence of irreducible information in networked computable systems that follow an information-sharing~(or communication) protocol based on contagion or infection models, as described in~\cite{Pastor-Satorras2001,Pastor-Satorras2001a,Pastor-Satorras2002}. As supported by these references, such spreading models, using the approach from complex networks, have been shown to be relevant to study epidemic and disease spreading, computer virus infections, or the spreading of polluting agents. Consequently, the study of such systems has helped on immunization strategies, epidemiology, or pollution control. 
	
	Nevertheless, instead of focusing on the pathological properties of such complex networks' contagion dynamics, we show a preliminary result on how such contagion dynamics might trigger an unlimited potential of optimization through diffusion. That is, diffusing the best solution (or the largest integer when one uses the Busy Beaver Game~\cite{Abrahao2017} as a toy model) through the network may trigger an unlimited increase of expected emergent algorithmic information of the nodes as the randomly generated population of computable systems (i.e., nodes) grows. Thus, we aim at mathematically investigating under which conditions this phenomenon is expected to happen. For this purpose, we use the theoretical framework for networked computable systems developed in~\cite{Abrahao2017} and the network model studied in~\cite{Pastor-Satorras2001,Pastor-Satorras2001a,Pastor-Satorras2002}.
	
	As a toy model, such theoretical approach to the study of emergence of com\-plex\-ity or information in networked computable systems may help understand and establish foundational properties on why an information dynamics within a system displaying synergistic or emergent behavior might be advantageous from a computational, evolutionary, or game-theoretical point of view. Moreover, as it is our goal to suggest in the present work, these phenomena may be also related to infection dynamics~(either from computer viruses or diseases) and scale-free networks.
	
	\section{Model}
	
	We define a model for randomly generated Turing machines that are networked with a scale-free degree distribution that obeys a power law of the form $ P(k) \sim 2m^2 \, k^{-3} $. The topology and construction of the networks are defined by a random process connecting new nodes under a probability distribution given by preferential attachment as in \cite{Barabasi1999,Barabasi2003}. That is, new nodes are more likely to establish connections to higher degree nodes. The diffusion or ``infection" scheme is ruled by the Susceptible-Infected-Susceptible model~(SIS), in which susceptible nodes have a constant probability $ \nu $ of being ``infected" by an ``infected" neighbor and an ``infected" node has a constant probability $ \delta $ of becoming ``cured". We also assume, as in \cite{Pastor-Satorras2001,Pastor-Satorras2001a,Pastor-Satorras2002}, that the prevalence of ``infected" nodes (i.e., the average density of ``infected" nodes) becomes stationary after sufficient time.
	
	However, nodes are now Turing machines that can send and receive information (partial outputs) as each node runs its computations until returning a final output. We have defined this population of randomly generated Turing machines and a more general mathematical model for networked computable systems which we have called as \emph{algorithmic networks}~\cite{Abrahao2016b,Abrahao2017}. This population plays the Busy Beaver Imitation Game (BBIG) in which each node always imitates the fittest neighbor only. However, differently from the one in~\cite{Abrahao2017}, we here present a variation on the information-sharing protocol. The difference in respect to this previous work comes from allowing nodes to become ``cured" (with rate $ \delta $). Besides, now nodes also get ``infected" with rate $ \nu $ --- which may have a different value from $1$ --- while in~\cite{Abrahao2017} one has that $ \nu = 1 $ always holds. While still playing a Busy Beaver Imitation Game, susceptible nodes follow a rule of imitating the neighbor that had output the largest integer, but they follow this rule with probability $ \nu $. Thus, the effective spreading rate $ \lambda = \nu / \delta $ defined in~\cite{Pastor-Satorras2001,Pastor-Satorras2001a,Pastor-Satorras2002} assumes a direct interpretation of the rate in which the imitation-of-the-fittest protocol~\cite{Abrahao2017} was applied on a node---and this is the reason why we are using the words ``infection" and ``cure" between quotation marks.
	
	\section{Results}
	
	Our proofs follow mainly from information theory, computability theory, and graph theory applied on a variation on the information-sharing~(communication) protocol of the model in~\cite{Abrahao2017}. From this model, we have proven results for general dynamic networks and for dynamic networks with small diameter, i.e., $\mathbf{O}(\log(N))$ compared to the network size~$N$. We have shown that there are topological conditions that trigger a phase transition in which eventually the algorithmic network $ \mathfrak{N}_{BB}  $ begins to produce an unlimited amount of bits of emergent algorithmic complexity (i.e., emergent algorithmic information). These conditions come from a positive trade-off between the average diffusion density and the number of cycles (or communication rounds). Therefore, the diffusion power of a dynamic network has proven to be paramount with the purpose of optimizing the average fitness/payoff of an algorithmic network that plays the Busy Beaver Imitation Game. Besides, this diffusion power may come either from the cover time~\cite{Costa2015a} or from a small diameter compared to the network size. 
	
	Therefore, our developed proofs\footnote{ The article presenting these and other results and complete and extended versions of these proofs is available at https://doi.org/10.5281/zenodo.1228505} for a network following the SIS diffusion model, as in~\cite{Pastor-Satorras2001,Pastor-Satorras2001a,Pastor-Satorras2002}, also stems from the main idea of combining an estimation of a lower bound for the average algorithmic complexity/information of a networked node and an estimation of an upper bound for the expected algorithmic complexity/information of an isolated node. The estimation of the latter still comes from the strong law of large numbers, Gibb's inequality, and algorithmic information theory applied on the randomly generated population.
	%
	However, now the estimation of the former comes from the SIS model with a stationary prevalence~(i.e., the average density of infected nodes), which gives directly this lower bound by the fact that the prevalence $ \rho \sim 2 \exp(- 1 / m \lambda) $ in~\cite{Pastor-Satorras2001,Pastor-Satorras2001a,Pastor-Satorras2002} becomes equal to the average diffusion density $ \tau_{\mathbf{E}} $ in~\cite{Abrahao2017}. 
	
	Thus, let condition $
	\tau_{\mathbf{E}} |_{t_0}^{c(N)} \sim 2 \exp(- 1 / m \lambda)  > \Omega( w, c(N) ) $
	\noindent holds where $c(N)$ is an upper bound for the time in which the network achieves stationary prevalence~(as in \cite{Pastor-Satorras2001a,Pastor-Satorras2002,Pastor-Satorras2001}). Additionally, we have that $ c(N) $ is a computable function of the network size~$N$\footnote{ In fact, as a function of $ N^C $, where $C$ is a constant that may have a value $ \leq 1 $.}, $t_0$ is the first time instant, $w$ is the network input available to every node at the beginning of the first cycle, and $\Omega(w,c(N))$ is the probability that a node (i.e., a Turing machine) halts, with $w$ as initial input, in every cycle until the last cycle $c(N)$. Note that $ \Omega(w,c(N))$ is a value that depends only on the chosen self-delimiting programming language in which one defines a universal Turing machine. Then, one can prove that the expected emergent algorithmic complexity/information of a node goes to infinity as the network size (i.e., the population size) $ N $ goes to infinity. 
	
	In other words, the average irreducible information that emerges when nodes are networked, compared with when they are isolated, is expected to always increase for large enough populations of randomly generated Turing machines that compute during $ \leq c(N) $ cycles (or communication rounds). Moreover, since it is an emergent phenomenon that arises depending only on the network/population size, one can show that such scale-free networks of randomly generated Turing machines are also expected to display, for large enough populations, a complexity/information phase transition, which we have called \emph{expected emergent open-endedness}~\cite{Abrahao2017}.  
	
	\section{Conclusions}
	
	First, we have defined a family of Barab\'{a}si-Albert scale-free networks with a power law on the degree distribution of the form $ P(k) \sim 2m^2 \, k^{-3} $ that follow a contagion (or infection) dynamics described by the Susceptible-Infected-Susceptible model as in~\cite{Pastor-Satorras2001,Pastor-Satorras2001a,Pastor-Satorras2002}. Their set of nodes are randomly generated Turing machines. Additionally, every node plays the SIS Busy Beaver Imitation Game, which ensures that each node always imitates the fittest neighbor whenever an ``infection'' would occur with probability $\nu$. Nodes also may return to its initial partial output---being ``cured''---with probability $\delta$.
	
	Then, we have shown that, for large enough values of $ m \, \lambda $, if the time for achieving a stationary prevalence of ``infected'' nodes $  \rho \sim 2 \exp(- 1 / m \lambda)  $ is upper bounded by a value given by a computable function of the network (or population) size $N$, then these networks of randomly generated Turing machines will have the property of expected emergent open-endedness for large enough network/population sizes. That is, the expected emergent algorithmic complexity/information of a node will eventually start to increase undefinitely as the population grows. Thus, this characterizes the complexity (or information) phase transition introduced in~\cite{Abrahao2017}.
	
	\section*{Acknowledgements}
	
	This work is partially funded by CAPES, CNPq, FAPESP, and FAPERJ.
	
	\bibliographystyle{sbc}
	\bibliography{1.1_ETC2018-2_FelipeKlausArtur}

\end{document}